\documentclass[preprint,12pt]{elsarticle}

\usepackage{amssymb}
\usepackage{graphicx}
\usepackage{dcolumn}
\usepackage{bm}
\usepackage{rotating}

\journal{Computer Physics Communications}

\newcommand{\kslash}{k\kern-1ex /}
\newcommand{\pslash}{p\kern-1ex /}
\newcommand{\qslash}{q\kern-1ex /}
\newcommand{\lslash}{l\kern-1ex /}
\newcommand{\sslash}{s\kern-1ex /}
\newcommand{\Dslash}{D\kern-1.2ex /}

\newcommand{\beqa}{\begin{eqnarray}}
\newcommand{\eeqa}{\end{eqnarray}}

\newcommand{\bd}{\begin{description}}
\newcommand{\ed}{\end{description}}

\newcommand{\ben}{\begin{eqnarray}}
\newcommand{\een}{\end{eqnarray}}
\newcommand{\nn}{\nonumber}
\def\lsim{\raise0.3ex\hbox{$<$\kern-0.75em\raise-1.1ex\hbox{$\sim$}}}
\def\gsim{\raise0.3ex\hbox{$>$\kern-0.75em\raise-1.1ex\hbox{$\sim$}}}
\def\simgt{\rlap{\lower 6.0 pt\hbox{$\mathchar \sim$}}\raise 2.5pt \hbox {$>$}}
\def\simlt{\rlap{\lower 6.0 pt\hbox{$\mathchar \sim$}}\raise 2.5pt \hbox {$<$}}

\newcommand{\csw}{{c_{\rm SW}}}

\begin{document}

\begin{frontmatter}



\title{Application of block Krylov subspace algorithms to 
the Wilson-Dirac equation
with multiple right-hand sides\\ in lattice QCD}


\author{T.~Sakurai$^{a}$, H.~Tadano$^{a,b}$, Y.~Kuramashi$^{c,b}$}

\address{
 $^a$Department of Computer Science, University of Tsukuba,\\ Tsukuba, Ibaraki 305-8573, Japan\\
 $^b$Center for Computational Sciences, University of Tsukuba,\\ Tsukuba, Ibaraki 305-8577, Japan\\
 $^c$Graduate School of Pure and Applied Sciences, University of Tsukuba,\\ Tsukuba, Ibaraki 305-8571, Japan\\
}

\begin{abstract}
It is well known that the block Krylov subspace solvers work
efficiently for some cases of the solution of differential
equations with multiple right-hand sides. In lattice QCD
calculation of physical quantities on a given configuration 
demands us to solve the Dirac equation with multiple
sources. 
We show that a new block Krylov subspace algorithm recently proposed by the
authors reduces the computational cost significantly
without loosing numerical accuracy for the solution of
the $O(a)$-improved Wilson-Dirac equation.
\end{abstract}


\end{frontmatter}


\thispagestyle{plain}

\section{Introduction}
\label{sec:intro}

In the last decade one of the primary issues in lattice QCD is
to reduce the dynamical up and down quark masses toward the physical values.
Most of our efforts have been devoted to reduce the
computational cost for the configuration generation with light quark masses.
Thanks to the algorithmic improvement together with 
rapid increase of the computational power we are now allowed 
to make a direct full QCD simulation
on the physical up and down quark masses\cite{pacscs_nf3,kura_latt08}.
On the other hand, we have been less concerned with 
the algorithmic improvement for the calculation of physical quantities:
its computational cost was negligibly smaller than 
that for the configuration generation until recently.
At present, however, the latter is drastically reduced to be 
comparable to or smaller than the former. 

The characteristic feature in the calculation of physical quantities
is the solution of the Dirac equation with multiple sources:
twelve in the simplest case and $O(10-100)$ for the stochastic technique.  
These are typical examples of the solution of differential equations
with multiple right-hand sides. For this type of equations it is known 
that the block Krylov subspace
solvers succeed in reducing the computational cost\cite{blockcg}.
In this article we study the application of the block 
BiCGSTAB algorithm\cite{blockbicgstab} 
and a new block Krylov subspace method proposed in Ref.~\cite{alg_tadano} 
to the $O(a)$-improved Wilson-Dirac
equation which is one of the popular fermion formulations
in current lattice QCD simulations. 
For simplicity our numerical test
is restricted on the quenched configuration with a fixed volume.
We investigate the quark mass dependence of the algorithmic efficiency 
in detail.  

This paper is organized as follows. In Sec.~\ref{sec:equation} we give 
the definition of the $O(a)$-improved Wilson-Dirac equation.
The algorithmic details are described in Sec.~\ref{sec:bicgstab}.
In Sec.~\ref{sec:test} we present the results of the numerical tests
after explaining the parameter choice and the machine specifications.
Our conclusions are summarized in Sec.~\ref{sec:conclusion}.

\section{Wilson-Dirac equation}
\label{sec:equation}

The lattice QCD is defined on a hypercubic four-dimensional lattice of
finite extent being expressed as 
$L_x\times L_y\times L_z\times L_t$ with
$L_{x,y,z}$ the three-dimensional spatial extent and $L_t$ 
the temporal one. 
The lattice spacing is set to unity for notational 
convenience. The fields are defined on the sites $n$ with periodic
boundary conditions. 

We define two types of fields on the lattice. One is the 
gauge field represented by $U_\mu(n,a,b)$ with $\mu=1,2,3,4$ and $a,b=1,2,3$
which is a $3\times 3$ SU(3) matrix assigned on each link.
The other is the quark field $q(n,\alpha,a)$ which resides on each site
carrying the Dirac index $\alpha=1,2,3,4$ and the color index $a=1,2,3$. 
The $O(a)$-improved Wilson-Dirac operator is written as
\ben
D_{\rm W}(n,\alpha,a;m,\beta,b)&=&
\delta_{n,m}\delta_{\alpha,\beta}\delta_{a,b}\nn\\
&&-\kappa\sum_{\mu=1,\dots ,4}\left[
\{\delta_{\alpha,\beta}-\gamma_\mu(\alpha,\beta)\}U_\mu(n,a,b)\delta_{m,n+{\hat \mu}}\right.\nn\\
&&\left.\;\;\;\;\;\;\;\;\;
+\{\delta_{\alpha,\beta}+\gamma_\mu(\alpha,\beta)\}U^\dagger_\mu(n-{\hat
\mu},a,b)\delta_{m,n-{\hat \mu}}
\right]\nn\\
&&-\kappa \cdot \csw\sum_{\mu,\nu=1,\dots ,4}\frac{i}{2}\sigma_{\mu\nu}(\alpha,\beta)
F_{\mu\nu}(n,a,b)\delta_{m,n},
\label{eq:wd-op}
\een
where ${\hat \mu}$ denotes the unit vector in the $\mu$ direction.
The coefficient $\csw$ is a parameter to be adjusted for 
the $O(a)$ improvement.
The Euclidean gamma matrices are defined in terms of 
the Minkowski ones in the Bjorken-Drell convention:
$\gamma_j=-i\gamma_{BD}^j$ $(j=1,2,3)$, 
$\gamma_4=\gamma_{BD}^0$,
$\gamma_5=\gamma_{BD}^5$ and 
$\sigma_{\mu\nu}=\frac{1}{2}[\gamma_\mu,\gamma_\nu]$.
The explicit representations for $\gamma_{1,2,3,4,5}$ are given by
\ben
\gamma_1&=&\left(\begin{array}{cccc}
             \hspace*{6mm}&\hspace*{6mm}&\hspace*{6mm}& -i \\
                  &            & -i         &\hspace*{6mm}  \\
                  &  i         &            &    \\
                i &            &            &     \\
            \end{array}\right),\\
\gamma_2&=&\left(\begin{array}{cccc}
              \hspace*{6mm}&\hspace*{6mm}&\hspace*{6mm}& -1 \\
                  &    &  1 &\hspace*{6mm}    \\
                  &  1 &    &    \\
               -1 &    &    &     \\
            \end{array}\right),\\
\gamma_3&=&\left(\begin{array}{cccc}
              \hspace*{6mm}&\hspace*{6mm}& -i &    \\
                  &    &    &  i \\
                i &    &\hspace*{6mm}&\hspace*{6mm}      \\
                  & -i &    &     \\
            \end{array}\right),\\
\gamma_4&=&\left(\begin{array}{cccc}
               1  &\hspace*{6mm}&\hspace*{6mm}&\hspace*{6mm}  \\
               \hspace*{6mm} &  1 &    &    \\
                  &    & -1 &    \\
                  &    &    & -1  \\
            \end{array}\right),\\
\gamma_5&=&\left(\begin{array}{cccc}
             \hspace*{6mm}&\hspace*{6mm}  &  1 &    \\
                  &    &    &  1 \\
                1 &    & \hspace*{6mm}&\hspace*{6mm}   \\
                  &  1 &    &     \\
            \end{array}\right),
\een
where we list only nonzero elements.
The field strength $F_{\mu\nu}$ in the last term of Eq.~(\ref{eq:wd-op})
is expressed as
\ben 
F_{\mu\nu}(n,a,b)&=&\frac{1}{4}\sum_{i=1}^{4}\frac{1}{2i}
\left(P_i(n,a,b)-P_i^\dagger(n,a,b)\right)
\een
with
\ben
P_1(n,a,b)&=&\sum_{c,d,e}U_\mu(n,a,c)U_\nu(n+{\hat \mu},c,d)
         U^\dagger_\nu(n+{\hat \nu},d,e)U^\dagger_\nu(n,e,b), \\
P_2(n,a,b)&=&\sum_{c,d,e}U_\nu(n,a,c)
         U^\dagger_\mu(n-{\hat \mu}+{\hat \nu},c,d)\nn\\ \vspace*{-2mm}
&&\;\;\;\;\;\;\times
         U^\dagger_\nu(n-{\hat \mu},d,e)U_\mu(n-{\hat \mu},e,b), \\
P_3(n,a,b)&=&\sum_{c,d,e}U^\dagger_\mu(n-{\hat \mu},a,c)
         U^\dagger_\nu(n-{\hat \mu}-{\hat \nu},c,d)\nn\\ \vspace*{-2mm}
&&\;\;\;\;\;\;\times
         U_\mu(n-{\hat \mu}-{\hat \nu},d,e)U_\nu(n-{\hat \nu},e,b), \\
P_4(n,a,b)&=&\sum_{c,d,e}U^\dagger_\nu(n-{\hat \nu},a,c)
         U_\nu(n-{\hat \nu},c,d)\nn\\ \vspace*{-2mm}
&&\;\;\;\;\;\;\times
         U_\nu(n+{\hat \mu}-{\hat \nu},d,e)U^\dagger_\mu(n,e,b).
\een
The $O(a)$-improved Wilson-Dirac operator defined by Eq.~(\ref{eq:wd-op})
is a complex non-Hermitian square matrix, where
only 51 out of $L_x\times L_y\times L_z\times L_t \times 3 \times 4$ 
entries in each row have nonzero values.
The matrix becomes fairly sparse in current numerical simulations 
with $L_{x,y,z,t}\sim O(10)$.

The calculation of physical quantities requires the solution of the
following linear equations:
\ben
\sum_{m,\beta,b}D_{\rm W}(n,\alpha,a;m,\beta,b)x^{(i)}(m,\beta,b)=s^{(i)}(n,\alpha,a),
\een
where $s^{(i)}$ represents the $i$-th source vector.
To illustrate the situation
we consider the calculation of the hadron two-point 
function with the point source at the origin as a simplest example.
In this case we need twelve source vectors expressed as
\ben
s^{(i)}(1,\alpha,a)=\left\{\begin{array}{cl}
            1  &  \;\; i=a+3(\alpha-1)\\
            0  &  \;\; {\rm otherwise}
\end{array}\right.
\een
with $a=1,2,3$ and $\alpha=1,2,3,4$.
Another good example is the stochastic technique which
usually requires $O(10-100)$ noise sources.
Although we can think of a lot of other interesting examples,
our numerical tests, which is explained later in Sec.~\ref{sec:test}, 
concentrate on the simplest one.

\section{Block Krylov subspace methods}
\label{sec:bicgstab}

Before explaining block Krylov subspace algorithms  
it should be better to reformulate the problem in a generalized form.
This help the readers easily understand the essence
avoiding any complex notations specific to lattice QCD.
  
Our interest exists in the solution of linear systems with the
multiple right-hand sides expressed as
\ben
AX=B,
\label{eq:linear_mrhs}
\een
where $A$ is an $N\times N$ complex sparse non-Hermitian matrix.
$X$ and $B$ are $N\times L$ complex rectangular matrices given by
\ben
X=\left(\bm{x}^{(1)},\dots,\bm{x}^{(i)},\dots,\bm{x}^{(L)} \right),\\
B=\left(\bm{b}^{(1)},\dots,\bm{b}^{(i)},\dots,\bm{b}^{(L)} \right).
\een
In the case of the Wilson-Dirac equation 
$N=L_x\times L_y\times L_z\times L_t \times 3 \times 4$
and $L$ is the number of the source vectors.

For a preparative purpose we first write down the well-known BiCGSTAB algorithm
for solving a single right-hand side linear system, 
where $\bm{x}=\bm{x}^{(1)}$ and $\bm{b}=\bm{b}^{(1)}$ in Eq.~(\ref{eq:linear_mrhs}):
\begin{quote}
$\bm{x}_0$ is an initial guess, \\
{\bf Compute} $\bm{r}_0=\bm{b}-A \bm{x}_0$, \\
{\bf Set} $\bm{p}_0=\bm{r}_0$, \\
{\bf Choose} $\tilde{\bm{r}}_0$ such that $(\tilde{\bm{r}}_0,\bm{r}_0)\ne 0$, \\
{\bf For} $k=0, 1, \dots, $ {\bf until}
$\Vert \bm{r}_k\Vert_2/\Vert \bm{b}\Vert_2\le \epsilon$ {\bf do:} \\
\begin{tabular}{rcl}
$\alpha_k$ &=& $(\tilde{\bm{r}}_0,\bm{r}_k)/(\tilde{\bm{r}}_0,A \bm{p}_k)$, \\ 
$\bm{t}_k$ &=& $\bm{r}_k-\alpha_k A \bm{p}_k$, \\
$\zeta_k$ &=& $(A \bm{t}_k, \bm{t}_k)/(A \bm{t}_k, A \bm{t}_k)$, \\
$\bm{x}_{k+1}$ &=& $\bm{x}_k+\alpha_k \bm{p}_k+\zeta_k \bm{t}_k$, \\
$\bm{r}_{k+1}$ &=& $\bm{t}_k-\zeta_k A \bm{t}_k$, \\
$\beta_k$ &=& $(\alpha_k/\zeta_k)\cdot (\tilde{\bm{r}}_0,\bm{r}_{k+1})/(\tilde{\bm{r}}_0,\bm{r}_k)$, \\
$\bm{p}_{k+1}$ &=& $\bm{r}_{k+1}+\beta_k (\bm{p}_k-\zeta_k A \bm{p}_k)$, 
\end{tabular}\\     
{\bf End for}.
\end{quote}
It is rather straightforward to extend the algorithm to the blocked
version for solving multiple right-hand sides linear system:
\begin{quote}
$X_0\in \mathbb{C}^{N\times L}$ is an initial guess, \\
{\bf Compute} $R_0=B-A X_0$, \\
{\bf Set} $P_0=R_0$, \\
{\bf Choose} $\tilde{R}_0 \in \mathbb{C}^{N\times L}$, \\
{\bf For} $k=0, 1, \dots,$ {\bf until}
$\max_i(\Vert \bm{r}_k^{(i)}\Vert_2
/\Vert \bm{b}^{(i)}\Vert_2)\le \epsilon$ {\bf do:} \\
\begin{tabular}{rcl}
$V_k$ &=& $AP_k$, \\
\multicolumn{3}{l}{{\bf Solve} $({\tilde{R}_0}^{\mathrm{H}} V_k)\alpha_k$ = ${\tilde{R}_0}^{\mathrm{H}}R_k$ 
                                           for $\alpha_k$,} \\ 
$T_k$ &=& $R_k-V_k \alpha_k$, \\
$Z_k$ &=& $AT_k$, \\
$\zeta_k$ &=& ${\rm Tr}\left(Z_k^{\mathrm{H}} T_k\right)
               /{\rm Tr}\left(Z_k^{\mathrm{H}} Z_k\right)$, \\
$X_{k+1}$ &=& $X_k+P_k \alpha_k +\zeta_k T_k$, \\
$R_{k+1}$ &=& $T_k-\zeta_k Z_k$, \\
\multicolumn{3}{l}{{\bf Solve} $({\tilde{R}_0}^{\mathrm{H}} V_k)\beta_k$ = $-{\tilde{R}_0}^{\mathrm{H}} Z_k$ 
                                           for $\beta_k$,} \\
$P_{k+1}$ &=& $R_{k+1}+(P_k-\zeta_k V_k)\beta_k$, 
\end{tabular}\\     
{\bf End for},
\end{quote}
where $R_k,P_k,T_k$ are $N\times L$ complex rectangular matrices and
$\alpha_k,\beta_k$ are $L\times L$ complex square ones.
At the $k$-th iteration in the block BiCGSTAB algorithm we find
\ben
X_k &\in & X_0 +{\cal K}_k^{\blacksquare}(A;R_0),\\
R_k &\in & {\cal K}_{k+1}^{\blacksquare}(A;R_0),
\een
where ${\cal K}_k^{\blacksquare}(A;R_0)$ is a block Krylov subspace defined by
\begin{displaymath}
 {\cal K}_k^{\blacksquare}(A;R_0) \equiv \left\{ \sum_{j=0}^{k-1}A^j R_0 \xi_j~;~\xi_j \in \mathbb{C}^{L \times L}\right\}.
\end{displaymath}
This yields an essential difference from the consecutive
application of the BiCGSTAB algorithm to $A\bm{x}^{(i)}=\bm{b}^{(i)}$ for
$i=1,\dots, L$: The blocked version searches the solution vectors
with the enlarged Krylov subspace.

In Ref.~\cite{alg_tadano} 
a new block Krylov subspace method is proposed. 
This method improves the numerical accuracy of the block BiCGSTAB
method where multiplication of the matrix $\alpha_k$
yields contaminations of the rounding error on the solution\cite{alg_tadano}.
The algorithm is as follows:

\begin{quote}
   $X_0 \in \mathbb{C}^{n \times L}$ is an initial guess, \\
   {\bf Compute} $R_0 = B - AX_0$, \\
   {\bf Set} $P_0 = R_0$ and $V_0 = W_0 = AR_0$, \\
   {\bf Choose} $\tilde{R}_0 \in \mathbb{C}^{n \times L}$, \\
   {\bf For} $k=0, 1, \dots, $ {\bf until}
$\max_i(\Vert \bm{r}_k^{(i)}\Vert_2
/\Vert \bm{b}^{(i)}\Vert_2)\le \epsilon$ {\bf do:} \\
   \begin{tabular}{rcl}
    \multicolumn{3}{l}{{\bf Solve} $({\tilde R}_0^{\mathrm{H}}V_k)\alpha_k = {\tilde R}_0^{\mathrm{H}}R_k$ for $\alpha_k$,} \\
    $\zeta_k$ &=& $\mathrm{Tr}\left[W_k^{\mathrm{H}}R_k\right] / \mathrm{Tr}\left[W_k^{\mathrm{H}}W_k\right]$, \\
    $S_k$ &=& $P_k - \zeta_k V_k$, \\
    $U_k$ &=& $S_k \alpha_k$, \\
    $Y_k$ &=& $AU_k$, \\
    $X_{k+1}$ &=& $X_k + \zeta_k R_k + U_k$, \\
    $R_{k+1}$ &=& $R_k - \zeta_k W_k - Y_k$, \\
    $W_{k+1}$ &=& $AR_{k+1}$, \\
    \multicolumn{3}{l}{{\bf Solve} $({\tilde R}_0^{\mathrm{H}}R_k)\gamma_k = {\tilde R}_0^{\mathrm{H}}R_{k+1} / \zeta_k$ for $\gamma_k$,} \\
    $P_{k+1}$ &=& $R_{k+1} + U_k \gamma_k$, \\
    $V_{k+1}$ &=& $W_{k+1} + Y_k \gamma_k$, \\
   \end{tabular}\\
   {\small \bf End}
\end{quote}

This algorithm is constructed to avoid the rounding error problem 
observed in the block BiCGSTAB method.

\section{Numerical tests}
\label{sec:test}

\subsection{Choice of parameters}
\label{subsec:params}

The $L$ dependence of the
efficiency of the block Krylov algorithms, in which we are most interested,  
should be investigated on single CPU avoiding contaminations
due to the communication overhead. 
The memory requirement forces the lattice size to be moderate. 
Our numerical tests are performed 
with samples of 10 statistically independent gauge field configurations 
on a $16^3\times 32$ lattice generated by the Iwasaki gauge action
at $\beta=2.575$ in quenched approximation, which was employed 
in Ref.~\cite{cppacs_nf2}. 
We solve the Wilson-Dirac equation with the local source at the
origin choosing four hopping parameters $\kappa=0.1359,0.1357,0.1355,0.1300$
with the improvement coefficient $\csw=1.345$.
The bare quark mass is given by $m_q=(1/\kappa-1/\kappa_c)/2$ with
$\kappa_c=0.136116(8)$, which increases as $\kappa$ decreases. 
In Table~\ref{tab:param} we list the pion mass in physical unit and 
the $m_\pi/m_\rho$ ratio
at each hopping parameter following Ref.~\cite{cppacs_nf2}.
Although the pion mass at $\kappa=0.1300$ is extremely large 
from a view point of physical interest, 
we employ it for a comparative purpose.  
The lattice spacing estimated by $m_\rho$ 
is $a=0.1130$ fm\cite{cppacs_nf2}.

\begin{table*}[htb]
\centering{
\caption{Pion mass and $m_\pi/m_\rho$ ratio at each hopping parameter.}
\label{tab:param}
\newcommand{\cc}[1]{\multicolumn{1}{c}{#1}}
\begin{tabular}{cccccc}
\hline
$\kappa$ & $\kappa_c=0.136116(8)$ & 0.1359 & 0.1357 & 0.1355 & 0.1300 \\
\hline
$m_\pi$ [MeV] &  0 & 221  & 307  & 375  & 1282 \\
$m_\pi/m_\rho$ & 0 & 0.28 & 0.39 & 0.46 & 0.87 \\
\hline
\end{tabular}
}
\end{table*}

\subsection{Test environment}
\label{subsec:implementaion}

Numerical tests are carried out on single node of T2K-Tsukuba 
which is a large-scale
cluster system 648 compute nodes providing
95.4Tflops of computing capability.
Each node consists of quad-socket, 2.3GHz Quad-Core AMD Opteron Model
8356 processors whose on-chip cache sizes are 64KBytes/core,
512KBytes/core, 2MB/chip for L1, L2, L3, respectively.  
Each processor has a direct connect memory interface to an 8GBytes 
DDR2-667 memory and three hypertransport links to connect other processors.
All the nodes in the system are connected through 
a full-bisectional fat-tree network 
consisting of four interconnection links of 8GBytes/sec aggregate
bandwidth with Infiniband.

\subsection{Results}
\label{subsec:results}

Table~\ref{tab:results_old} shows the $L$ dependence of 
the computational cost to solve 
the Wilson-Dirac equation with the block BiCGSTAB algorithm 
imposing rather stringent tolerance 
$\epsilon=10^{-14}$ for the relative residual. We present averaged values
over 10 configuration samples for the number of iteration
and the execution time. The true residual, which is defined by
$\max_i \Vert A x^{(i)}-b^{(i)}\Vert_2/\Vert b^{(i)}\Vert_2$,  
is evaluated after the relative residual reaches the tolerance. 
The maximum and minimum values among 10 configuration samples are listed.  
Although the computational cost divided by $L$ is considerably
reduced as $L$ increases at $\kappa=0.1359,0.1357,0.1355$,
there exist two concerns: One is the discrepancy between the relative residual
and the true one which is enhanced as $L$ increases. The other is
the convergence failure of the relative residual, which is found 
for 4 samples out of 10 at $\kappa=0.1359$ with $L=4$.
This flaw is also observed at $\kappa=0.1355$ and 0.1357 
once we go beyond $L=4$.   

\begin{figure*}[h]
\vspace{13mm}
\begin{center}
\includegraphics[width=110mm,angle=0]{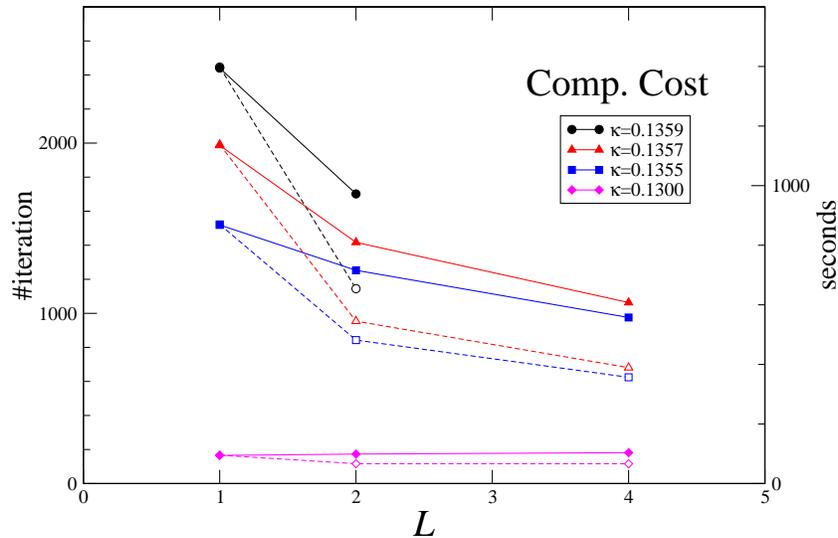}  
\end{center}
\vspace{-.5cm}
\caption{$L$ dependence of the number of iteration (solid) and
the execution time divided by $L$ (open) for the new method. All the
 results are averaged over 10 configuration samples.}
\label{fig:cost}
\end{figure*}

\begin{figure*}[h]
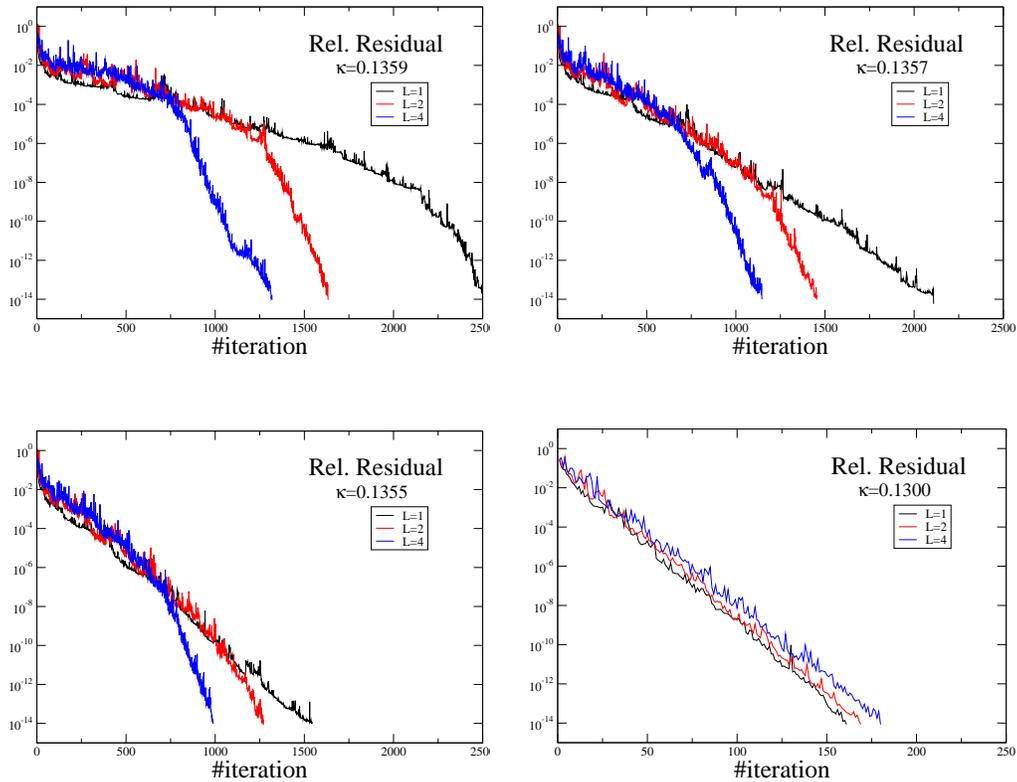

\vspace{13mm}
\begin{center}
\begin{tabular}{cc}
\includegraphics[width=65mm,angle=0]{figs/residuals/res_k01359.eps}
&
\includegraphics[width=65mm,angle=0]{figs/residuals/res_k01357.eps}\\
\vspace*{3mm}& \\
\includegraphics[width=65mm,angle=0]{figs/residuals/res_k01355.eps}
&
\includegraphics[width=65mm,angle=0]{figs/residuals/res_k01300.eps}
\end{tabular}
\end{center}
\vspace{-.5cm}
\caption{Representative behaviors of the relative residual as a
function of the number of iteration. All the measurements are performed on
the same configuration.}
\label{fig:residual}
\end{figure*}

In order to remove the discrepancy of the relative residual 
and true one, we proposed a new algorithm 
whose details are presented in Ref.~\cite{alg_tadano}.
The effectiveness of the new method is
observed in Table~\ref{tab:results_new}, where 
the deviation between the true and the relative 
residuals essentially vanishes. 
Figure~\ref{fig:cost} plots the number of iteration 
and the execution time divided $L$ as a function of 
$L$. We observe two important features.  One 
is the acceleration of the cost reduction at the light quark masses where
the physical interest exists: At $\kappa=0.1357$ the number of 
iteration is almost cut in half from $L=1$ to 4, which should be compared to
the case of $\kappa=0.1300$.
The other is that the executional time
divided by $L$ decreases faster than the number of iteration 
as $L$ increases: The former is reduced by more than 60\%
from $L=1$ to 4 at $\kappa=0.1357$. 
This fact demonstrates the efficiency of the cache-aware 
implementation for the matrix multiplication on the multiple vectors.
In Fig.~\ref{fig:residual} we show a representative behavior of the
relative residual as a function of the number of iteration at each
hopping parameter.
We observe little $L$ dependence up to 700 iterations, beyond which 
the convergence speed is accelerated as $L$ increases 
at $\kappa=0.1359,0.1357,0.1355$.

\begin{table*}[htb]
\centering{
\caption{$L$ dependence of the number of iteration and the execution time
to solve the Wilson-Dirac equations with the block BiCGSTAB method.
True residual is evaluated after the relative
residual reaches the tolerance $\epsilon=10^{-14}$.}
\label{tab:results_old}
\newcommand{\cc}[1]{\multicolumn{1}{c}{#1}}
\begin{tabular}{rrrrrrr}
\hline
$\kappa$ & $L$ & \#iteration & time& time/$L$&
 \multicolumn{2}{c}{true residual} \\
& & &  \multicolumn{1}{c}{ [sec] } & \multicolumn{1}{c}{ [sec] } &  \multicolumn{1}{c}{max} & \multicolumn{1}{c}{min} \\
\hline
0.1359 & 1 & 2432.6 & 1439.3 & 1439.3 & $1.57\times 10^{-14}$ & $7.99\times 10^{-15}$\\
       & 2 & 1696.8 & 1316.0 & 658.0 & $2.91\times 10^{-11}$ & $2.36\times 10^{-13}$\\
       & 4 & $-$ & $-$ & $-$ & $-$ & $-$\\
\hline
0.1357 & 1 & 1999.0 & 1162.4 & 1162.4 & $1.08\times 10^{-14}$ & $5.98\times 10^{-15}$\\
       & 2 & 1410.1 & 1092.4 & 546.2 & $2.62\times 10^{-12}$ & $3.75\times 10^{-14}$\\
       & 4 & 1100.0 & 1633.3 & 408.3 & $1.36\times 10^{-11}$ & $6.33\times 10^{-13}$\\
\hline
0.1355 & 1 & 1518.5 &  884.4 & 884.4 & $1.09\times 10^{-14}$ & $5.25\times 10^{-15}$\\
       & 2 & 1264.2 &  979.6 & 489.8 & $2.55\times 10^{-12}$ & $1.49\times 10^{-14}$\\
       & 4 &  961.7 & 1430.0 & 357.5 & $1.30\times 10^{-11}$ & $2.30\times 10^{-13}$\\
\hline
0.1300 & 1 & 165.3 &  96.4 & 96.4 & $9.29\times 10^{-15}$ & $3.68\times 10^{-15}$\\
       & 2 & 172.8 & 134.3 & 67.2 & $1.57\times 10^{-14}$ & $6.14\times 10^{-15}$\\
       & 4 & 181.0 & 272.7 & 68.2 & $3.42\times 10^{-13}$ & $7.99\times 10^{-15}$\\
\hline
\end{tabular}
}
\end{table*}

\begin{table*}[htb]
\centering{
\caption{$L$ dependence of the number of iteration and the execution time
to solve the Wilson-Dirac equations with the new method.
True residual is evaluated after the relative
residual reaches the tolerance $\epsilon=10^{-14}$.}
\label{tab:results_new}
\newcommand{\cc}[1]{\multicolumn{1}{c}{#1}}
\begin{tabular}{rrrrrrr}
\hline
$\kappa$ & $L$ & \#iteration & time & time/$L$ &
 \multicolumn{2}{c}{true residual} \\
& & & \multicolumn{1}{c}{ [sec] }  & \multicolumn{1}{c}{ [sec] }  &  \multicolumn{1}{c}{max} & \multicolumn{1}{c}{min} \\
\hline
0.1359 & 1 & 2440.3 & 1398.2 & 1398.2 & $1.53\times 10^{-14}$ & $8.57\times 10^{-15}$\\
       & 2 & 1701.5 & 1308.4 & 654.2 & $2.92\times 10^{-13}$ & $6.65\times 10^{-15}$\\
       & 4 & $-$ & $-$ & $-$ & $-$ & $-$\\
\hline
0.1357 & 1 & 1986.6 & 1138.1 & 1138.1 & $1.14\times 10^{-14}$ & $6.81\times 10^{-15}$\\
       & 2 & 1417.0 & 1090.1 & 545.1 & $1.08\times 10^{-14}$ & $8.51\times 10^{-15}$\\
       & 4 & 1063.6 & 1556.6 & 389.2 & $1.51\times 10^{-14}$ & $6.72\times 10^{-15}$\\
\hline
0.1355 & 1 & 1519.0 &  870.0 & 870.0 & $1.18\times 10^{-14}$ & $9.03\times 10^{-15}$\\
       & 2 & 1252.9 &  962.9 & 481.5 & $1.08\times 10^{-14}$ & $7.68\times 10^{-15}$\\
       & 4 &  975.7 & 1427.1 & 356.8 & $1.29\times 10^{-14}$ & $7.23\times 10^{-15}$\\
\hline
0.1300 & 1 & 165.4 &  95.3 & 95.3 & $9.34\times 10^{-15}$ & $4.02\times 10^{-15}$\\
       & 2 & 173.1 & 133.6 & 66.8 & $9.79\times 10^{-15}$ & $5.15\times 10^{-15}$\\
       & 4 & 181.7 & 266.1 & 66.5 & $9.10\times 10^{-15}$ & $6.79\times 10^{-15}$\\
\hline
\end{tabular}
}
\end{table*}

\section{Conclusions}
\label{sec:conclusion}

In this paper we present the first example for the efficiency 
of the block Krylov subspace methods in lattice QCD 
to solve the $O(a)$-improved Wilson-Dirac equation 
with multiple local sources. 
We find remarkable cost reductions for the light quark masses.
Roughly speaking, the solver performance normalized by $L$ is doubled
by increasing $L$ from one to four at the light quark masses.
The block Krylov subspace methods have two advantageous points.
Firstly we can easily implement the method 
by extending the conventional solver for a single right hand side. 
The deviation between the relative residual and the true one 
observed in the block BiCGSTAB algorithm is 
successfully removed by the new algorithm proposed 
in Ref.~\cite{alg_tadano}.
Although our numerical tests are carried out on a single CPU,
it is obvious that there is no difficulty in parallelization.
Secondly the multiplication of the Wilson-Dirac matrix on the multiple
vectors allow us an effective use of the cache, where the link
variables can be retained during the operation.
The cost reduction is achieved by not only the algorithmic efficiency
but also the implementation technique. 
One concern about the methods is that the increase of $L$ makes 
difficult the convergence of the relative residuals at light
quark masses.
We are now investigating its origin and possible improvements.

\section*{Acknowledgments}
Numerical calculations for the present work have been carried out
on the T2K-Tsukuba computer at the University of Tsukuba.
This work is supported in part by Grants-in-Aid for Scientific Research
from the Ministry of Education, Culture, Sports, Science and Technology
(Nos.
20800009,   
18540250    
).



\end{document}